\documentclass[journal]{IEEEtran}

\usepackage{color}
\usepackage{amsmath}
\usepackage{amsthm}
\usepackage{bm}
\usepackage{amsfonts}
\usepackage{graphicx}
\usepackage{multirow}
\usepackage{url}
\usepackage{hyperref}
\usepackage[flushleft]{threeparttable}
\usepackage{ifpdf} 
\usepackage{cite}
\usepackage{array}
\usepackage[normalem]{ulem}
\usepackage[table]{xcolor}
\usepackage{cuted}

\usepackage{diagbox}
\usepackage{stfloats}
\usepackage{enumerate}
\usepackage{booktabs}
\usepackage{float}
\usepackage{arydshln}
\usepackage{mathtools}
\usepackage{amssymb}

\begin{document}

\title{Chance-Constrained OPF in Droop-Controlled Microgrids with Power Flow Routers}

\author{Tianlun~Chen,
		David J.~Hill,
		Yue~Song,
        and~Albert Y.S. Lam

\vspace{-0.5cm}
}

\markboth{}
{Shell \MakeLowercase{\textit{et al.}}: Bare Demo of IEEEtran.cls for IEEE Journals}

\maketitle

\begin{abstract}
High penetration of renewable generation poses great challenge to power system operation due to its uncertain nature. In droop-controlled microgrids, the voltage volatility induced by renewable uncertainties is aggravated by the high droop gains. This paper proposes a chance-constrained optimal power flow (CC-OPF) problem with power flow routers (PFRs) to better regulate the voltage profile in microgrids. PFR refer to a general type of network-side controller that brings more flexibility to the power network. Comparing with the normal CC-OPF that relies on power injection flexibility only, the proposed model introduces a new dimension of control from power network to enhance system performance under renewable uncertainties. Since the inclusion of PFRs complicates the problem and makes common solvers no longer apply directly, we design an iterative solution algorithm. For the subproblem in each iteration, chance constraints are transformed into equivalent deterministic ones via sensitivity analysis, so that the subproblem can be efficiently solved by the convex relaxation method. The proposed method is verified on the modified IEEE 33-bus system and the results show that PFRs make a significant contribution to mitigating the voltage volatility and make the system operate in a more economic and secure way.
\end{abstract}


\IEEEpeerreviewmaketitle

\section{Introduction} \label{sec:introduction}
Microgrids (MGs) refer to low-voltage or medium-voltage power networks integrated with renewable distributed generations (DGs), loads and other control devices \cite{farrokhabadi2019microgrid}, which can operate in either grid-connected mode or islanded mode. MGs have drawn much attention in the recent decades due to its flexibility for arbitrary configurations with different sizes and functionalities. However, the high penetration of renewable DGs has brought many challenges to MG operators in handling the economic and security issues. Commonly, islanded MGs adopt droop control schemes for autonomous power sharing between the dispatchable DGs \cite{guerrero2010hierarchical}. High droop gains are usually used for better transient response and proper power sharing \cite{barklund2008energy}. However, this feature makes the voltages and frequency even more sensitive to power injection changes and thus leads to highly volatile voltages under renewable uncertainties \cite{song2017distributed}. 

Optimal power flow (OPF) is a fundamental tool for voltage regulation and economic dispatch in power system operations. Traditionally, OPF is formulated as a deterministic problem, which optimizes an objection function (e.g., generation cost) subjected to operational constraints such as voltage and line flow limits \cite{wood2013power}. However, the deterministic OPF is not sufficient to ensure an economic and secure operation with the presence of uncertainties, especially for droop-controlled MGs where the voltage volatility (i.e., the degree of voltage variance under uncertainties) is further aggravated by the high droop gains.    
Recently, chance-constrained OPF (CC-OPF) has become a powerful tool for addressing the challenges brought by high penetration of renewables \cite{zhang2011chance, wang2011chance}. Different from the deterministic OPF, CC-OPF replaces the hard constraints by chance constraints to guarantee that the probabilities of constraint violations under uncertain disturbances are kept within pre-defined values. 
Existing studies have shown the effectiveness of CC-OPF in accommodating renewable energy in low voltage systems. For example, the reactive power support from DG inverters are utilized to mitigate voltage variations under renewable uncertainties by chance constrained optimization \cite{lopez2017two, li2018distribution, nazir2018two}. 
Controllable loads are also utilized through chance constrained framework to achieve a more economic generation and reserve scheduling \cite{vrakopoulou2017chance} or reduce the power losses while maintaining an acceptable voltage profile \cite{hassan2018optimal}. In another line of work, multi-period optimization \cite{dall2017chance} and model predictive control \cite{jiang2018stochastic} are designed with chance constraints using battery energy storage systems (BESSs) to hedge the negative impacts of renewable uncertainties.


The above CC-OPF models mainly rely on the power injection flexibility provided by node-side devices. 
The potential of network flexibility in CC-OPF has not been much exploited yet. Nowadays power systems have increasing network flexibility enabled by advanced power electronic devices where power flow router (PFR) is a representative example. PFR was first proposed in \cite{7436826} as a general type of controller installed at lines that makes the power network more flexible. PFR introduces a new mechanism into system control that tunes the routing of power injections rather than the conventional node-side flexibility which modifies the power injections. 
In our previous work \cite{chen2020reducing}, PFR was introduced as an effective way to reduce the BESS capacity required for accommodating renewable energy. The obtained results in \cite{chen2020reducing} indicate that the network flexibility makes great contribution to the voltage regulation in the corresponding multi-period OPF problem. It inspires us to apply PFRs in CC-OPF problems to address the voltage volatility caused by renewable uncertainties.

The contributions of this paper are twofold:
\begin{enumerate}
\item To the authors' knowledge, this is the first formulation of AC CC-OPF considering MG droop characteristics and network flexibility. 
We combine the CC-OPF model with the droop characteristics and PFRs; PFRs introduce a new dimension of control which is shown to make significant contribution to the voltage volatility reduction. It leads to a more economic and secure operating status against renewable uncertainties.

\item We design an iterative algorithm which is tailored for the proposed optimization problem. For the subproblem in each iteration, the chance constraints are reformulated into deterministic equivalents by sensitivity analysis, which enables the subproblem to fully utilize the efficiency of existing AC-OPF algorithms, e.g., semidefinite programming (SDP) relaxation. 
\end{enumerate}
The reminder of the paper is organized as follows. Section \ref{sec:problem} provides the system modelling and the optimization problem formulation. Section \ref{sec:method} describes the solution methodology. In Section \ref{sec:simulation}, case studies are presented to evaluate the performance of the proposed model and algorithm. Finally, conclusions and future work are given in Section  \ref{sec:conclusion}.


\section{Problem Formulation} \label{sec:problem}
Consider an islanded microgrid with the set of buses $\mathcal{N}:= \{1,2,...,n\}$ and the set of lines $\mathcal{E} \subseteq \mathcal{N} \times \mathcal{N}$. Each bus may connect a dispatchable DG, a renewable DG and a load. For bus $i$, the active and reactive power generations of the dispatchable DG are denoted as $P_{G_i}$ and $Q_{G_i}$; the active and reactive power generations of the renewable DG is denoted as $P_{W_i}$ and $Q_{W_i}$; the active and reactive power loads are denoted as $P_{L_i}$ and $Q_{L_i}$. For bus $k$ without DG generations or loads, the respective notations $P_{G_k}$, $Q_{G_k}$, $P_{W_k}$, $Q_{W_k}$, $P_{L_k}$, $Q_{L_k}$ are always zero. Also, denote $V_i$ and $\theta_i$ as the voltage magnitude and voltage angle at bus $i$. 
A line connecting bus $i$ and bus $j$ is denoted by an unordered pair $(i,j) \in \mathcal{E}$ and some lines are installed with PFRs. 
In the following, we detail the system models and the optimization problem formulation.

\subsection{Droop-Controlled Dispatchable DGs}
A dispatchable DG refers to a DG unit whose output can be adjusted by the operators. We denote the set of buses with dispatchable DG generations as $\mathcal{N_G} \in \mathcal{N}$.
The dispatchable DGs are assumed to adopt the conventional P-$\omega$ and Q-V droop control \cite{song2017distributed}, which is expressed as 
\begin{subequations}
\begin{align} 
\omega &= \omega^{\ast} - K_{p_i}(P_{G_i} - P_{G_i}^{\ast}), i\in \mathcal{N_G} \label{p-droop}\\
V_i   &=  V^{\ast}_i - K_{q_i}(Q_{G_i} - Q_{G_i}^{\ast}),  i\in \mathcal{N_G}\label{q-droop}
\end{align}
\end{subequations}
where $\omega$ is the angular frequency of the system; $\omega^{\ast}$ and $V^{\ast}_i$ are the set points of frequency and voltage magnitude; $K_{p_i}$ and $K_{q_i}$ are the frequency and voltage droop gains; $P_{G_i}^{\ast}$ and $Q_{G_i}^{\ast}$ are the set points of active and reactive power generation.

\subsection{Renewable DGs and Uncertainty Modelling}
A renewable DG is considered as a non-dispatchable source. We assume the renewable DGs follow the maximum power point tracking mode which introduce uncertainties into the power network. The active power generation of renewable DG at bus $i$ is modelled as sum of the forecasted value ${P}_{W_i}^f$ and forecast error $\xi_i$
\begin{equation}
P_{W_i}(\bm{\xi}) = {P}_{W_i}^f + \xi_i,
\end{equation} 
where $\bm{\xi} = [\xi_i] \in \mathbb{R}^{n}$ is a vector of forecast errors which follows a multivariate distribution featured by zero mean and known covariance matrix $\Sigma \in \mathbb{R}^{n\times n}$. For bus $k$ without renewable DG generation, the corresponding $k$-th row and column of $\Sigma$ is set to zero.   
Moreover, we assume a constant power factor $\lambda_i$ so that the reactive power generation at bus $i$ follows the active power generation: 
\begin{equation} \label{Reactive Generation}
{Q}_{W_i}(\bm{\xi}) = \lambda_i({P}_{W_i}^f + \xi_i),
\end{equation}
where $\lambda_i = \tan{\phi}_i $ determines the reactive power control of renewable DG at bus $i$. Similar to above, for bus $k$ without renewable DG generation, $\lambda_k$ is always zero. 
For simplicity, we consider the renewable DGs as the only source of power injection uncertainties and our model can be easily extended to include load uncertainties.

\subsection{Power Flow Equations with Power Flow Routers}
The AC power flow equations are adopted for accurately describing the behaviours under renewable uncertainties.
For a normal line $(i,j)$ without PFRs, the active and reactive branch power flows $P_{ij}$ and $Q_{ij}$ are given as
\begin{subequations}
\begin{align} 
&P_{ij} = g_{ij}(V_i^2 - V_iV_j\cos \theta_{ij}) -  b_{ij}V_iV_j\sin \theta_{ij}, \label{BFM-1} \\
&Q_{ij} =  -b_{ij}(V_i^2 - V_iV_j\cos \theta_{ij}) -  g_{ij}V_iV_j\sin \theta_{ij},  \label{BFM-2}
\end{align}
\end{subequations}
where $y_{ij} = g_{ij} + \mathrm{j}b_{ij}$ represents the admittance of line $(i,j)$; the notation $\theta_{ij}$ is the short for $\theta_{i}- \theta_{j}$.

\begin{figure}[t] 
\includegraphics[width=\linewidth]{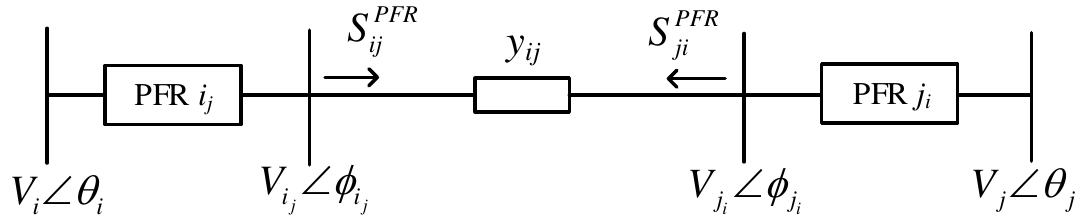} \vspace{-0.5cm} 
\caption{A diagram for a line with PFRs. } \vspace{-0.5cm} 
\label{fig:PFR_branch}
\end{figure}

PFRs are installed at some lines to bring network flexibility to the system and enlarge the feasible region of OPF problems \cite{7436826, chen2020robust}. 
The diagram of a line with PFRs is shown in Fig. \ref{fig:PFR_branch}, where PFRs refer to a pair of series voltage regulators (tuning both magnitude and phase) installed at two terminals of the line. A typical implementation of PFR is power electronic transformer, which has been used in both high-voltage and low-voltage systems.  It is natural to introduce PFRs into microgrids as power electronics devices are more and more ubiquitous. 
Literature \cite{7436826,chen2020robust, chen2020reducing} have investigated the value of PFRs in different OPF formulations for loadability enhancement, cost reduction and voltage regulation. Further, it will be seen later that PFRs can benefit CC-OPF for voltage regulation by tuning both the mean values and standard deviations of voltages, while power injection dispatch is only effective in tuning the mean values.

The branch power flow with PFRs is presented as follows.
As shown in Fig. \ref{fig:PFR_branch}, $V_{i_j}\angle \phi_{i_j}$ and $V_{j_i}\angle \phi_{j_i}$ refer to the complex voltages of the secondary sides of the PFRs. The relations between  $V_{i_j}\angle \phi_{i_j}$, $V_{j_i}\angle \phi_{j_i}$ and $V_i$, $V_j$ are given by
\begin{subequations}
\begin{align} 
V_{i_j}\angle \phi_{i_j} &= T_{i_j}^*V_{i}\angle (\theta_i + \beta_{i_j}^*) \label{eq: PFR voltage1} \\
V_{j_i}\angle \phi_{j_i} &= T_{j_i}^*V_{j}\angle (\theta_j + \beta_{j_i}^*) \label{eq: PFR voltage2}
\end{align}
\end{subequations}
where $T_{i_j}^*$ and $T_{j_i}^*$ are the tap ratio set points of PFR $i_j$ and $j_i$; $\beta_{i_j}^*$ and $\beta_{j_i}^*$ are the phase shift set points of PFR $i_j$ and $j_i$. 
The PFRs are assumed to have no conversion losses \cite{7436826}. Thus, for a line $(i,j)$ with PFRs, the active and reactive branch power flows are expressed as
\begin{subequations}
\allowdisplaybreaks[4]
\begin{align}
P^{PFR}_{ij} &= g_{ij}(V_{i_j}^2 - V_{i_j}V_{j_i}\cos(\phi_{i_j}-\phi_{j_i})) \nonumber \\
&\quad \quad    -  b_{ij}V_{i_j}V_{j_i}\sin (\phi_{i_j}-\phi_{j_i}) \nonumber\\
&  = g_{ij}(T_{i_j}^{*2}V_i^2 - T_{i_j}^*T_{j_i}^*V_iV_j\cos(\theta_{ij} + \beta_{i_jj_i}^*)) \nonumber\\
& \quad\quad - b_{ij}T_{i_j}^*T_{j_i}^*V_iV_j\sin(\theta_{ij} +\beta_{i_jj_i}^* ) \label{BFM-PFR-1}\\
Q^{PFR}_{ij} &= -b_{ij}(V_{i_j}^2 - V_{i_j}V_{j_i}\cos(\phi_{i_j}-\phi_{j_i})) \nonumber \\
&\quad \quad-  g_{ij}V_{i_j}V_{j_i}\sin(\phi_{i_j}-\phi_{j_i})   \nonumber \\
&  = -b_{ij}(T_{i_j}^{*2}V_i^2 - T_{i_j}^*T_{j_i}^*V_iV_j\cos(\theta_{ij} + \beta_{i_jj_i}^*)) \nonumber\\
& \quad\quad - g_{ij}T_{i_j}^*T_{j_i}^*V_iV_j\sin(\theta_{ij} +\beta_{i_jj_i}^* )  \label{BFM-PFR-2}
\end{align}
\end{subequations}
where the notation $\beta_{i_jj_i}^* $ is the short for  $\beta_{i_j}^* - \beta_{j_i}^*$.
For simplicity, branch power flows without PFRs \eqref{BFM-1}-\eqref{BFM-2} can be transformed into  \eqref{BFM-PFR-1}-\eqref{BFM-PFR-2} by setting $T_{i_j}^* = T_{j_i}^* = 1$ and $\beta_{i_j}^* = \beta_{j_i}^* = 0$. Thus, the active and reactive power balance at each bus $i$ can be given by the unified expression below 
\begin{subequations}
\begin{align}
 P_{G_i} + P_{W_i}(\bm{\xi}) - P_{L_i} 
      &=  \sum_{(i,j)\in \mathcal{E}} P^{PFR}_{ij}(\bm{V},\bm{\theta},\bm{T}^*,\bm{\beta}^*)  \label{BIM-1}  \\      
Q_{G_i} + Q_{W_i}(\bm{\xi}) - Q_{L_i}
      &=  \sum_{(i,j)\in \mathcal{E}} Q^{PFR}_{ij}(\bm{V},\bm{\theta},\bm{T}^*,\bm{\beta}^*)  \label{BIM-2}
\end{align}
\end{subequations}
where $ P_{L_i}$ and $Q_{L_i}$ are the active and reactive power loads at bus $i$;
$\bm{V} \in \mathbb{R}^n $ and $\bm{\theta} \in \mathbb{R}^n $ stack $V_i$ and  $\theta_i$, respectively; $\bm{T}^*:= (T_{i_j}^*, T_{j_i}^*, (i,j) \in \mathcal{E})$, $\bm{\beta}^*:= (\beta_{i_j}^*, \beta_{j_i}^*, (i,j) \in \mathcal{E})$. We also define the vectors $\bm{P}_G,\bm{Q}_G, \bm{P}_W(\bm{\xi}),\bm{Q}_W(\bm{\xi}) \in \mathbb{R}^n $ stacking $P_{G_i}$, $Q_{G_i}, P_{W_i}(\bm{\xi}), Q_{W_i}(\bm{\xi})$, respectively.

\subsection{Optimization Problem Formulation}
As stated in the previous section, given the set points $\{\bm{P}_G^*, \bm{Q}_G^*, \omega^*,  \bm{V}^*, \bm{T}^*, \bm{\beta}^*\}$ and renewable power generation $\bm{P}_W$, $\bm{Q}_W$, we can determine the values of $\{\bm{P}_{G}, \bm{Q}_{G}, \bm{V}, \bm{\theta}, \omega\}$ based on \eqref{p-droop}-\eqref{q-droop} and \eqref{BFM-PFR-1}-\eqref{BIM-2}. Since $\bm{P}_W$ and $\bm{Q}_W$ are functions of uncertainty $\bm{\xi}$, $\{ \bm{P}_{G}, \bm{Q}_{G}, \bm{V}, \bm{\theta}, \omega \}$ are not only subject to one possible realization of $\bm{\xi}$ but to a variety of renewable power realizations. Thus, they can also be expressed as the implicit functions of $\bm{\xi}$, say $\{ \bm{P}_{G}(\bm{\xi}), \bm{Q}_{G}(\bm{\xi}), \bm{V}(\bm{\xi}), \bm{\theta}(\bm{\xi}), \omega(\bm{\xi}) \}$, which describe the system responses to the uncertainty realizations.
Hence, we can formulate the CC-OPF with PFRs (CC-OPF-PFR) as follows
\begin{subequations} \label{CCOPF}
\allowdisplaybreaks[4]
\begin{align} 
\min \quad &  \mathbb{E}[ \sum_{i \in \mathcal{N_G}}c_{2i}{P_{G_i}(\bm{\xi})}^2 + c_{1i}P_{G_i}(\bm{\xi}) + c_{0i}], \label{COP-1} \\
s.t. \quad & \eqref{BFM-PFR-1}-\eqref{BFM-PFR-2}, \label{COP-18} \\
&P_{G_i}(\bm{\xi}) + P_{W_i}(\bm{\xi}) - P_{L_i} =   \nonumber\\
&\sum_{(i,j)\in \mathcal{E}} P^{PFR}_{ij}(\bm{V}(\bm{\xi}),\bm{\theta}(\bm{\xi}),\bm{T}^*,\bm{\beta}^*) ,\forall i \in \mathcal{N}  \label{COP-19}  \\      
&Q_{G_i}(\bm{\xi}) + Q_{W_i}(\bm{\xi}) - Q_{L_i} =   \nonumber\\
& \sum_{(i,j)\in \mathcal{E}} Q^{PFR}_{ij}(\bm{V}(\bm{\xi}),\bm{\theta}(\bm{\xi}),\bm{T}^*,\bm{\beta}^*) ,\forall i \in \mathcal{N}  \label{COP-20} \\
& \omega(\bm{\xi}) = \omega^{\ast} - K_{p_i}(P_{G_i}(\bm{\xi}) - P_{G_i}^{\ast}), \forall i \in \mathcal{N_G} \label{COP-21}\\
&V_i(\bm{\xi}) =  V^{\ast}_i - K_{q_i}(Q_{G_i}(\bm{\xi}) - Q_{G_i}^{\ast}), \forall i \in \mathcal{N_G} \label{COP-22} \\
&\mathbb{P}({P}_{G_i}(\bm{\xi}) \leq {P}_{G_i}^{\text{max}} ) \geq 1 - \epsilon_P, \forall i \in \mathcal{N_G} \label{COP-3} \\
&\mathbb{P}({P}_{G_i}(\bm{\xi}) \geq {P}_{G_i}^{\text{min}} ) \geq 1 - \epsilon_P, \forall i \in \mathcal{N_G} \label{COP-4} \\
&\mathbb{P}({Q}_{G_i}(\bm{\xi}) \leq {Q}_{G_i}^{\text{max}} ) \geq 1 - \epsilon_Q,  \forall i \in \mathcal{N_G} \label{COP-5} \\
&\mathbb{P}({Q}_{G_i}(\bm{\xi}) \geq {Q}_{G_i}^{\text{min}} ) \geq 1 - \epsilon_Q, \forall i \in \mathcal{N_G} \label{COP-6} \\
&\mathbb{P}({V}_{i}(\bm{\xi}) \leq {V}_{i}^{\text{max}} ) \geq 1 - \epsilon_V, \forall i \in \mathcal{N} \label{COP-7} \\
&\mathbb{P}({V}_{i}(\bm{\xi}) \geq {V}_{i}^{\text{min}} ) \geq 1 - \epsilon_V, \forall i \in \mathcal{N} \label{COP-8} \\
&\mathbb{P}({\omega}(\bm{\xi}) \leq {\omega}^{\text{max}} ) \geq 1 - \epsilon_{\omega},   \label{COP-9} \\
&\mathbb{P}({\omega}(\bm{\xi}) \geq {\omega}^{\text{min}} ) \geq 1 - \epsilon_{\omega},   \label{COP-10} \\
&{P}_{G_i}^{\text{min}} \leq P_{G_i}^{\ast} \leq {P}_{G_i}^{\text{max}}, \forall i \in \mathcal{N_G} \label{COP-11} \\
&{Q}_{G_i}^{\text{min}} \leq Q_{G_i}^{\ast} \leq {Q}_{G_i}^{\text{max}}, \forall i \in \mathcal{N_G}  \label{COP-12} \\
&{V}_{i}^{\text{min}} \leq V_{i}^{\ast} \leq {V}_{i}^{\text{max}},	 \forall i \in \mathcal{N}\label{COP-13} \\
&{\omega}^{\text{min}} \leq  \omega^{\ast} \leq {\omega}^{\text{max}},    \label{COP-14} \\
&{\gamma}_{i_j}^{\text{min}} \leq T_{i_j}^* \leq {\gamma}_{i_j}^{\text{max}}, \forall (i,j)\in \mathcal{E} \label{COP-15} \\
&{\beta}_{i_j}^{\text{min}} \leq \beta_{i_j}^* \leq {\beta}_{i_j}^{\text{max}}, \forall (i,j)\in \mathcal{E}  \label{COP-16}\\ 
&\theta_1 = 0.  \label{COP-17}
\end{align}
\end{subequations}

In this formulation, the objective function \eqref{COP-1} is to minimize the expected generation cost of the dispatchable DGs, where $c_{2i}, c_{1i}, c_{0i}$ are the cost coefficients.
Constraints \eqref{COP-18}--\eqref{COP-22} describe the power balance equations with PFRs and droop characteristics.
Constraints \eqref{COP-3}--\eqref{COP-6} are the chance constraints for active power and reactive power generation of dispatchable DGs. 
Constraints \eqref{COP-7}--\eqref{COP-10} are the chance constraints for voltage magnitudes and system frequency.
The chance constraint restricts the feasible region of OPF to a desired confidence region. In other words, it ensures the probability of constraint violation under any realization of uncertainties to be lower than a pre-specified level $\epsilon$.
Constraints \eqref{COP-11}--\eqref{COP-14} give the limits for set points of power outputs of dispatchable DGs, voltage magnitudes and system frequency, respectively.
Constraints \eqref{COP-15}--\eqref{COP-16} represent the upper and lower limits for tuning variables of PFRs.
We denote bus 1 as the reference bus and make $\theta_1$ fixed to zero in \eqref{COP-17}.

One major advantage of the CC-OPF-PFR model is that the generation dispatch is coordinated with PFR tuning to regulate voltages under renewable uncertainties. According to \cite{barklund2008energy}, the values of droop gains are usually large in MGs. However, this setting may result in volatile voltages and even make CC-OPF infeasible. On the other hand, it will be seen that the employment of PFRs considerably reduces voltage volatility.

Note that the set points $\{\bm{P}_G^*, \bm{Q}_G^*, \omega^*,  \bm{V}^*, \bm{T}^*, \bm{\beta}^*\}$ are determined by problem \eqref{CCOPF} and remain constant under renewable generation fluctuations. Hence, the chance constraints \eqref{COP-7}--\eqref{COP-10} refer to the system responses with the fixed set points under different renewable power scenarios and need to be satisfied with the prescribed violation probabilities.


\section{Solution Methodology} \label{sec:method}
The difficulty of solving problem \eqref{CCOPF} is that the non-linearity of power flow equations introduces significant challenges to quantify the system behaviours under uncertainties. This is different from the linear power flow model by which we can explicitly model the system responses to renewable uncertainties. It also explains why most literature consider linear  power flow model \cite{vrakopoulou2013probabilistic,bienstock2014chance,lubin2015robust, dall2017chance} by which chance constraint can be reformulated to an analytical form so that the problem can be more easily solved. However, the linear power flow models cannot accurately describe voltage behaviours under uncertainties, which necessitates the adoption of non-linear AC power flow equations. 
To ensure the tractability of the AC CC-OPF problem, literature have proposed several methods. 
For instance, authors in \cite{8017474} develop a method to get the approximate analytical form of chance constraints by iteratively linearizing around the operating point. On the other hand, authors in \cite{8060613} propose a SDP relaxation of AC CC-OPF and introduce piecewise affine approximation to achieve the tractability of chance constraints. But this method assumes fixed network parameters, which does not apply to our model where both network parameters and power injections are variables.
Moreover, authors in \cite{nazir2018two} adopt the scenario approach to address the chance constraints but it suffers from the computational burden of power flow calculations for a large number of scenarios.  

Similar to \cite{8017474}, we linearize the AC power flow equations around a given operating point to model the system responses under uncertainties. This linearization is based on the fact that the renewable forecast techniques have been developed with satisfactory performance so that the forecast errors are quite small. Thus, the chance constraints can be transformed into analytical forms and this also leads to a more tractable reformulation of problem \eqref{CCOPF}. Based on the chance constraint reformulation via linearization, we design an iterative algorithm to solve the optimization problem which will be detailed in the following subsections. 

\subsection{Chance Constraint Reformulation by Power Flow Linearization}
As stated in Section \ref{sec:problem}, we can obtain a operating point $(\bm{V}, \bm{\theta}, \omega)$ under the forecasted renewable power scenario $\bm{\xi} = \bm{0}$ and certain set points $\{\bm{P}_G^*, \bm{Q}_G^*, \omega^*,  \bm{V}^*, \bm{T}^*, \bm{\beta}^*\}$. This operating point satisfies the power flow equations \eqref{COP-18}-\eqref{COP-22} which are rewritten into a compact form for simplicity
\begin{align}
\bm{f}(\bm{P}_{W}(\bm{0}), \bm{Q}_{W}(\bm{0}), &\bm{V}(\bm{0}), \bm{\theta}(\bm{0}), \omega(\bm{0})) =\bm{0}.\label{FOP} 
\end{align}
Linearizing \eqref{FOP} around the given operating point gives the relation between the change of renewable power generations (i.e., $\bm{\xi}$ with small values) and the change of voltages and frequency
\begin{align} \label{linearization1}
\left[
\begin{array}{c} \Delta \bm{P}_W \\ \Delta\bm{Q}_W \\ {0} \end{array}
\right] &=
 \bm{J}_{PF} 
\left[
\begin{array}{c} \Delta\bm{ \theta}  \\ \Delta\bm{ V} \\ {\Delta \omega} \end{array}
\right], 
\end{align} 
where the notation $\Delta$ denotes the deviation from the given operating point and we define $\Delta\bm{ P}_W = [\Delta P_{W_i}]$, $\Delta\bm{ Q}_W = [\Delta Q_{W_i}]$, $\Delta\bm{ V} = [\Delta V_i]$, $\Delta\bm{ \theta} = [\Delta \theta_i]$ $\in \mathbb{R}^n$. Also, we denote $\bm{J}_{PF} \in \mathbb{R}^{(2n+1)\times(2n+1)}$ as the Jacobian matrix which can be expressed as
\begin{align} \label{linearization2}
\bm{J}_{PF}
=
\left[
\begin{array}{c  c  c}
  \bm{A} & \bm{B}  &\bm{S}_P \\
 \bm{C} & \bm{D} + \bm{S}_Q  & \bm{0} \\
  \bm{e}_1 & \bm{0}  & \bm{0}
\end{array}   
\right].
\end{align}
where the last row of $\bm{J_}{PF}$ represents the reference bus angle is fixed to zero; $\bm{S}_P = [S_{P_i}] \in \mathbb{R}^n$ is defined such that $S_{P_i} = K_{P_i}^{-1}, i \in \mathcal{N_G}$ and $S_{P_i} = 0, i \in \mathcal{N}\setminus\mathcal{N_G}$. For simplicity, a diagonal matrix $\bm{H} = \text{diag}\{h_1, h_2,... h_p\} \in \mathbb{R}^{p\times p}$ is denoted as $\bm{H} = \text{diag}\{h_i\} \in \mathbb{R}^{p\times p}$. Then, we define $\bm{S}_Q$ as 
$\bm{S}_Q = \text{diag}\{ S_{Q_i} \}  \in \mathbb{R}^{n\times n}$ such that $S_{Q_i} = K_{Q_i}^{-1}, i \in \mathcal{N_G}$ and $S_{Q_i} = 0, i \in \mathcal{N}\setminus\mathcal{N_G}$.
The sub-matrix $\bm{e}_1$ represents the vector with the first entry being one and the other entries being zero.
The sub-matrices $\bm{A}, \bm{B}, \bm{C}, \bm{D} \in \mathbb{R}^{n \times n}$ are derived from the power flow equations  \eqref{p-droop}--\eqref{q-droop} and \eqref{BIM-1}--\eqref{BIM-2}. The detailed expressions of $(\bm{A})_{ij}, (\bm{B})_{ij}, (\bm{C})_{ij}, (\bm{D})_{ij}$ are as follows
\begin{equation}
\allowdisplaybreaks[4]
\begin{split}
       &(\bm{A})_{ij} =   \left\{ \begin{array}{ l l }
\sum_{k=1}^n ( g_{ik}T_{i_k}^*T_{k_i}^*V_iV_k\sin(\theta_{ik}+ \beta_{i_kk_i}^*) - \\
\qquad\quad b_{ik}T_{i_k}^*T_{k_i}^*V_iV_k\cos(\theta_{ik}+ \beta_{i_kk_i}^*) ), i = j \\
g_{ij}T_{i_j}^*T_{j_i}^*V_iV_j\sin (\theta_{ij}+ \beta_{i_jj_i}^*) +  \\
\qquad\quad b_{ij}T_{i_j}^*T_{j_i}^*V_iV_j\cos(\theta_{ij}+ \beta_{i_jj_i}^*), i \neq j \\
                         \end{array}
        \right. \\
		&(\bm{B})_{ij} =   \left\{ \begin{array}{ l l }
 \sum_{k=1}^n ( g_{ik}(2T_{i_k}^*V_i - T_{k_i}^*V_k\cos (\theta_{ik}+ \beta_{i_kk_i}^*)) - \\
\qquad\quad b_{ik}T_{k_i}^*V_k\sin (\theta_{ik}+ \beta_{i_kk_i}^*)  ) , i = j \\
g_{ij}(-T_{i_j}^*V_i\cos (\theta_{ij}+ \beta_{i_jj_i}^*)) - \\
\qquad\quad b_{ij}T_{i_j}^*V_i\sin (\theta_{ij}+ \beta_{i_jj_i}^*), i \neq j \\
                         \end{array}
        \right. \\
        &(\bm{C})_{ij} =   \left\{ \begin{array}{ l l }
\sum_{k=1}^n ( -b_{ik}T_{i_k}^*T_{k_i}^*V_iV_k\sin (\theta_{ik}+ \beta_{i_kk_i}^*) - \\
\qquad\quad g_{ik}T_{i_k}^*T_{k_i}^*V_iV_k\cos (\theta_{ik}+ \beta_{i_kk_i}^*) ), i = j \\
-b_{ij}T_{i_j}^*T_{j_i}^*V_iV_j\sin (\theta_{ij}+ \beta_{i_jj_i}^*) + \\
\qquad\quad g_{ij}T_{i_j}^*T_{j_i}^*V_iV_j\cos (\theta_{ij}+ \beta_{i_jj_i}^*), i \neq j \\
                         \end{array}
        \right. \\
		&(\bm{D})_{ij} =   \left\{ \begin{array}{ l l }
\sum_{k=1}^n ( -b_{ik}(2T_{i_k}^*V_i - T_{k_i}^*V_k\cos (\theta_{ik}+ \beta_{i_kk_i}^*)) - \\
\qquad\quad g_{ik}T_{k_i}^*V_k\sin (\theta_{ik}+ \beta_{i_kk_i}^*)  ),i = j \\
-b_{ij}(-T_{i_j}^*V_i\cos (\theta_{ij}+ \beta_{i_jj_i}^*)) - \\
\qquad\quad g_{ij}T_{i_j}^*V_i\sin (\theta_{ij}+ \beta_{i_jj_i}^*), i \neq j. \\
                         \end{array}
        \right. 
\end{split}
\end{equation}
From the above equations we derive the expressions for the changes of system frequency and voltages with respect to renewable power fluctuations:
\begin{align} \label{linearization3}
\left[
\begin{array}{c} \Delta\bm{ \theta}  \\ \Delta\bm{ V} \\ {\Delta \omega}  \end{array}
\right] = 
 \bm{J}_{inv} 
\left[
\begin{array}{c}\Delta \bm{ P}_W \\ \Delta\bm{ Q}_W \\ {0} \end{array}
\right].
\end{align}
Substituting \eqref{Reactive Generation} to \eqref{linearization3} and dividing $\bm{J}_{inv}$ to sub-matrices give
\begin{align} 
\left[
\begin{array}{c} \Delta\bm{ \theta}  \\ \Delta\bm{ V} \\ {\Delta \omega}  \end{array}
\right] = 
\left[
\begin{array}{c c c}
  \bm{J}_{inv}^{11}  & \bm{J}_{inv}^{12} & \bm{J}_{inv}^{13} \\
 \bm{J}_{inv}^{21} & \bm{J}_{inv}^{22}  & \bm{J}_{inv}^{23} \\
  \bm{J}_{inv}^{31} & \bm{J}_{inv}^{32}  & \bm{J}_{inv}^{33}
\end{array}   
\right] 
\left[
\begin{array}{c} \Delta \bm{P}_W \\ \bm{\lambda}\Delta\bm{ P}_W \\ {0} \end{array}
\right],
\end{align} 
where $\bm{\lambda} = \text{diag}(\lambda_i) \in \mathbb{R}^{n \times n} $; $\bm{J}_{inv}$ is the inversion of Jacobian matrix $\bm{J}_{PF}$ and it can be divided into sub-matrices $\bm{J}_{inv}^{11}, \bm{J}_{inv}^{12}, \bm{J}_{inv}^{21}, \bm{J}_{inv}^{22} \in \mathbb{R}^{n \times n}$, $\bm{J}_{inv}^{31}, \bm{J}_{inv}^{32} \in \mathbb{R}^{1 \times n}$, $\bm{J}_{inv}^{13}, \bm{J}_{inv}^{23} \in \mathbb{R}^{n}$ and $ \bm{J}_{inv}^{33} \in \mathbb{R}$. The above equations allow us to obtain the following expressions for $\Delta\bm{ V}$ and $\Delta \omega$
\begin{equation} 
\Delta\bm{ V} = \bm{L}_V \Delta \bm{P}_W, \quad
\Delta\omega = \bm{L}_{\omega} \Delta \bm{P}_W, 
\end{equation} 
where $\bm{L}_{V} \in \mathbb{R}^{n\times n}$ and $\bm{L}_{\omega}\in \mathbb{R}^{1\times n}$ represent the sensitivities of voltage magnitudes and frequency to  $\Delta \bm{P}_W$, respectively; $\bm{L}_{V}$ and $\bm{L}_{\omega}$ can be calculated by
\begin{equation}
\bm{L}_{V} = \bm{J}_{inv}^{21} + \bm{\lambda}\bm{J}_{inv}^{22},\; \bm{L}_{\omega} = \bm{J}_{inv}^{31} + \bm{\lambda}\bm{J}_{inv}^{32}.
\end{equation}
Based on the chain rule and the droop characteristics we have
\begin{align} 
\Delta\bm{ P}_G &= \frac{\partial \bm{P}_G}{ \partial\omega} \Delta \omega = \bm{S}_P \bm{ L}_{\omega} \Delta\bm{ P}_W  \triangleq  \bm{L}_P \Delta\bm{P}_W,\\
\Delta\bm{Q}_G &= \frac{\partial \bm{Q}_G}{ \partial\bm{V}} \Delta \bm{V} = \bm{S}_Q \bm{ L}_{\bm{V}} \Delta\bm{ P}_W \triangleq  \bm{L}_Q \Delta\bm{P}_W,
\end{align} 
where $\bm{L}_{P} \in \mathbb{R}^{n \times n} $ and $\bm{L}_{Q}\in \mathbb{R}^{n \times n}$ represent the sensitivities of active and reactive power generation of dispatchable DGs. 

Based on above equations, the sensitivity matrices $\bm{L}_{P},\bm{L}_{Q}, \bm{L}_{V}, \bm{L}_{\omega}$ can be derived, which characterize the system responses under small uncertainties $\bm{\xi}$. Note that the sensitivity matrices are implicit and nonlinear functions of decision variables $\{\bm{V}, \bm{\theta}, \bm{T}^*, \bm{\beta}^* \}$ since they are given by the inversion of $\bm{J}_{PF}$.

\subsection{Analytical Reformulation of Chance Constraints}
Next, we further derive an analytical reformulation of the chance constraints based on the assumption that the renewable uncertainty $\bm{\xi}$ follows a multivariate normal distribution, with zero mean value and known covariance matrix $\Sigma$. 
First, using the sensitivity matrices $\bm{L}_{P},\bm{L}_{Q}, \bm{L}_{V}, \bm{L}_{\omega}$, chance constraints \eqref{COP-3}--\eqref{COP-10} can be approximated as linear functions of renewable uncertainties 
\begingroup
\allowdisplaybreaks[4]
\begin{align}
&\mathbb{P}({P}_{G_i}(\bm{0}) + \bm{L}_{P(i,\cdot)}\bm{\xi} \leq {P}_{G_i}^{\text{max}} ) \geq 1 - \epsilon_P,   &\forall i \in \mathcal{N_G} \label{LDR-1} \\
&\mathbb{P}({P}_{G_i}(\bm{0}) - \bm{L}_{P(i,\cdot)}\bm{\xi} \geq {P}_{G_i}^{\text{min}} ) \geq 1 - \epsilon_P,  &\forall i \in \mathcal{N_G} \label{LDR-2} \\
&\mathbb{P}({Q}_{G_i}(\bm{0}) + \bm{L}_{Q(i,\cdot)}\bm{\xi} \leq {Q}_{G_i}^{\text{max}} ) \geq 1 - \epsilon_Q,   &\forall i \in \mathcal{N_G} \label{LDR-3} \\
&\mathbb{P}({Q}_{G_i}(\bm{0}) - \bm{L}_{Q(i,\cdot)}\bm{\xi} \geq {Q}_{G_i}^{\text{min}} ) \geq 1 - \epsilon_Q,  &\forall i \in \mathcal{N_G} \label{LDR-4} \\
&\mathbb{P}(V_i(\bm{0}) + \bm{L}_{V(i,\cdot)}\bm{\xi} \leq {V}_{i}^{\text{max}} ) \geq 1 - \epsilon_V, 	  &\forall i \in \mathcal{N} \label{LDR-5} \\
&\mathbb{P}(V_i(\bm{0}) - \bm{L}_{V(i,\cdot)}\bm{\xi} \geq {V}_{i}^{\text{min}} ) \geq 1 - \epsilon_V,    &\forall i \in \mathcal{N} \label{LDR-6} \\
&\mathbb{P}(\omega(\bm{0}) + \bm{L}_{\omega(1,\cdot)}\bm{\xi} \leq {\omega}^{\text{max}} ) \geq 1 - \epsilon_{\omega},    \label{LDR-7} \\
&\mathbb{P}(\omega(\bm{0}) - \bm{L}_{\omega(1,\cdot)}\bm{\xi} \geq {\omega}^{\text{min}} ) \geq 1 - \epsilon_{\omega},   \label{LDR-8}
\end{align}
\endgroup
where the notations with subscript $(i,\cdot)$ refer to the i-th row of the respective matrices.
Then, based on the property of normal distribution, \eqref{LDR-1}--\eqref{LDR-8} are equivalent to the deterministic constraints as
\begingroup
\allowdisplaybreaks[4]
\begin{align}
&{P}_{G_i}(\bm{0}) + \kappa_{P}\text{Dev}\{P_{G_i}(\bm{\xi})\} \leq {P}_{G_i}^{\text{max}} ,   &\forall i \in \mathcal{N_G}  \label{LCC-1} \\
&{P}_{G_i}(\bm{0}) - \kappa_{P}\text{Dev}\{P_{G_i}(\bm{\xi})\} \geq {P}_{G_i}^{\text{min}} ,  &\forall i \in \mathcal{N_G} \label{LCC-2} \\
&{Q}_{G_i}(\bm{0}) + \kappa_{Q}\text{Dev}\{Q_{G_i}(\bm{\xi})\} \leq {Q}_{G_i}^{\text{max}} ,   &\forall i \in \mathcal{N_G} \label{LCC-3} \\
&{Q}_{G_i}(\bm{0}) - \kappa_{Q}\text{Dev}\{Q_{G_i}(\bm{\xi})\} \geq {Q}_{G_i}^{\text{min}} ,  &\forall i \in \mathcal{N_G} \label{LCC-4} \\
&V_i(\bm{0}) + \kappa_{V}\text{Dev}\{V_i(\bm{\xi})\} \leq {V}_{i}^{\text{max}} , 	 & \forall i \in \mathcal{N} \label{LCC-5} \\
&V_i(\bm{0}) - \kappa_{V}\text{Dev}\{V_i(\bm{\xi})\} \geq {V}_{i}^{\text{min}}  ,    &\forall i \in \mathcal{N} \label{LCC-6} \\
&\omega(\bm{0}) + \kappa_{\omega}\text{Dev}\{\omega(\bm{\xi})\} \leq {\omega}^{\text{max}}  ,    \label{LCC-7} \\
&\omega(\bm{0}) - \kappa_{\omega}\text{Dev}\{\omega(\bm{\xi})\} \geq {\omega}^{\text{min}}  ,   \label{LCC-8}
\end{align}
\endgroup
where $\kappa_{P} = \Phi^{-1}(1-\epsilon_P)$ denotes the inverse cumulative distribution function of normal distribution evaluated at $(1-\epsilon_P)$; similar interpretations apply to $\kappa_{Q},\kappa_{V},\kappa_{\omega}$; $\text{Dev}\{V_i(\bm{\xi})\}$ is the standard deviation of voltage magnitude at bus $i$
\begin{equation}
\text{Dev}\{V_i(\bm{\xi})\}  = \sqrt{\bm{L}_{V(i,\cdot)} \Sigma (\bm{L}_{V(i,\cdot)})^T}.
\end{equation}
Similar interpretations apply to $\text{Dev}\{P_{G_i}(\bm{\xi})\}$, $\text{Dev}\{Q_{G_i}(\bm{\xi})\}$, $\text{Dev}\{\omega(\bm{\xi})\}$ for generation outputs and system frequency. From \eqref{LCC-1}--\eqref{LCC-8}, we observe that the effect of renewable uncertainties in the chance constraint is equivalent to reducing the upper bound or increasing the lower bound by a uncertainty margin in the deterministic constraint. 
This transform of chance constraints is extendable to other types of distributions by using the same expression for standard deviation and a different $\Phi$ corresponding to that distribution.


\subsection{Reformulation of CC-OPF-PFR}
Similar to the transform of chance constraints, the objective function \eqref{COP-1} can be re-expressed as
\begin{align}
&\mathbb{E}[ \sum_{i \in \mathcal{N_G}}c_{2i}{P_{G_i}(\bm{\xi})}^2 + c_{1i}P_{G_i}(\bm{\xi}) + c_{0i}]  = \nonumber \\
&\sum_{i \in \mathcal{N_G}}c_{2i} [({P_{G_i}(0)}^2 + \bm{L}_P\Sigma (\bm{L}_P)^T] + c_{1i}P_{G_i}(0) + c_{0i}.
\end{align}
If the power loss is neglected, $\bm{L}_P$ is purely determined by droop gains and hence remains constant. This is indeed the case in low voltage systems such as MGs, where the power loss is very small and negligible. Thus, the term $\bm{L}_P\Sigma (\bm{L}_P)^T$ can be approximately regarded as a constant and excluded from the objective function. The objective function can be simplified into the total cost of dispatchable DGs under the forecasted renewable power scenario $\bm{\xi} = 0$. Then, we denote $\bar{P}_{G_i} = P_{G_i}(0)$ and the CC-OPF-PFR problem \eqref{CCOPF} can be reformulated as follows 
\begin{equation} \label{CCOPF-r}
\begin{split}
\text{minimize} \quad  &\sum_{i \in \mathcal{N_G}}(c_{2i}{\bar{P}_{G_i}}^2 + c_{1i}\bar{P}_{G_i} + c_{0i}) ,  \\
\text{subject to}  \quad &\eqref{COP-18}-\eqref{COP-22}\; \text{for} \; \bm{\xi} = 0, \\
&\eqref{COP-11} - \eqref{COP-17},  \eqref{LCC-1} - \eqref{LCC-8}.
\end{split}
\end{equation}
As shown above, the standard deviation terms $\text{Dev}\{\cdot\}$ in \eqref{LCC-1}--\eqref{LCC-8} are determined by $\{\bm{V}, \bm{\theta}, \bm{T}^*, \bm{\beta}^* \}$. Obviously, $\bm{T}^*$ and $\bm{\beta}^*$ contribute more to the change of standard deviation terms $\text{Dev}\{\cdot\}$ than $\bm{V}$ and $\bm{\theta}$ because the per-unit values of voltage magnitudes are close to one and the voltage angles are close to zero in low voltage systems. To satisfy the more stringent constraints \eqref{LCC-1}-\eqref{LCC-8}, traditional CC-OPF mainly relies on tuning the operating point by directly adjusting the power injections. However, from the perspective of network flexibility, we can expect that the PFRs not only tune the operating point but also reduce the standard deviations of voltage magnitudes (i.e., lower voltage volatility), which will be seen in the case studies.    

If the sensitivity matrices $\bm{L}_{P},\bm{L}_{Q}, \bm{L}_{V}, \bm{L}_{\omega}$ keep constant in \eqref{CCOPF-r}, \eqref{CCOPF-r} is a deterministic OPF problem and can be solved by SDP relaxation which is a well-established efficient solver \cite{low2014convex}. However, as stated above, the sensitivity matrices are implicit and nonlinear functions of decision variables $\{\bm{V}, \bm{\theta}, \bm{T}^*, \bm{\beta}^* \}$ which make the problem \eqref{CCOPF-r} difficult to solve. To tackle this issue, the sensitivity matrices need to be determined at a given operating point and fixed in problem \eqref{CCOPF-r}. In this way, the analytical formulations of chance constraints in \eqref{CCOPF-r} describe the case of the given operating point rather than the optimal solution to \eqref{CCOPF-r}. In other words, the optimal solution to \eqref{CCOPF-r} may not satisfy the chance constraints since there is a mismatch between the standard deviation terms $\text{Dev}\{\cdot\}$ in \eqref{CCOPF-r} and their actual values at the optimal solution. Therefore, an iterative algorithm is required to gradually eliminate such kind of mismatch.

\subsection{Solution Algorithm}
For the convenience of algorithm statement, let us denote the uncertainty margin of $V_i$ at the $k$-th iteration by ${\Omega}_{V_i}^k = \Phi^{-1}(1-\epsilon_V)\sqrt{\bm{L}^k_{V(i,\cdot)} \Sigma (\bm{L}^k_{V(i,\cdot)})^T} $. Similar interpretations apply to ${\Omega}^k_{P_i}$, ${\Omega}^k_{Q_i}$, ${\Omega}^k_{\omega}$ and we define the vectors $\bm{\Omega}^k_P,\bm{\Omega}^k_Q, \bm{\Omega}^k_V \in \mathbb{R}^n $ stacking ${\Omega}^k_{P_i},{\Omega}^k_{Q_i},{\Omega}^k_{V_i}$, respectively.
We summarize the iterative algorithm as follows. 
\begin{enumerate}[Step 1:]
\item Initialize the sensitivity matrices $\bm{L}_P^0, \bm{L}_Q^0, \bm{L}_V^0, {L}_{\omega}^0$ as zero matrices and the iteration count $k = 0$. 
\item Using modified SDP relaxation to solve the problem \eqref{CCOPF-r} with specified sensitivity matrices $\bm{L}_P^{k}, \bm{L}_Q^{k}, \bm{L}_V^{k}, {L}_{\omega}^{k}$ and obtain the solution $\bm{x}^{k+1} = $ $(\bm{V}^{k+1}$, $\bm{\theta}^{k+1}$, ${\omega}^{k+1}$, $\bm{P}_G^{*k}$, $\bm{Q}_G^{*k+1}$, $\omega^{*k+1}$,  $\bm{V}^{*k+1}$, $\bm{T}^{*k+1}$, $\bm{\beta}^{*k+1}$).
\item Calculate the sensitivity matrices $\bm{L}_P^{k+1}, \allowbreak \bm{L}_Q^{k+1}, \bm{L}_V^{k+1}, \allowbreak \bm{L}_{\theta}^{k+1}$  and margins $\bm{\Omega}^k = (\bm{\Omega}^k_P,\allowbreak\bm{\Omega}^k_Q,\allowbreak \bm{\Omega}^k_V, \allowbreak\Omega_{\omega}^k)$ at $\bm{x}^{k+1}$. 
Evaluate the maximum difference of $\bm{\Omega}$ between current iteration and the last iteration: $\Delta \bm{\Omega} = ||\bm{\Omega}^{k+1} - \bm{\Omega}^{k}||_{\infty}$. 
\item Check convergence: If $\Delta \bm{\Omega} \leq \delta$, stop. Otherwise, set $k = k+1$ and go back to Step 2.
\end{enumerate}


The iterative algorithm converges when the maximum deviation of uncertainty margins is smaller than $\delta$, which has a pre-defined value, e.g., $10^{-5}$. 
The physical meaning of the converged solution is as follows:
this solution takes the minimal cost to satisfy all the constraints, including those chance constraints in the form of \eqref{LCC-1}--\eqref{LCC-8} where the sensitivity matrices are obtained by the linearization around this solution.

By applying the above iterative algorithm, the basic structure of AC-OPF can be retained so that a modified SDP-based convex relaxation method on OPF with PFRs \cite{7436826,chen2020robust} can be fully utilized. It should be noted that SDP relaxation of the subproblem in each iteration is not an equivalent transformation. But literature have shown the exactness is commonly satisfied so that the solution obtained by SDP relaxation is equivalent to the solution to the original subproblem, which is also the situation in our case study.
Furthermore, the SDP relaxation can be replaced by any other fast OPF solvers. This solution algorithm is always easy to implement because the subproblem in each iteration has a simple form which excludes the computational complexity introduced by renewable uncertainties.


\begin{figure}[!t]
 \includegraphics[width=3.4in]{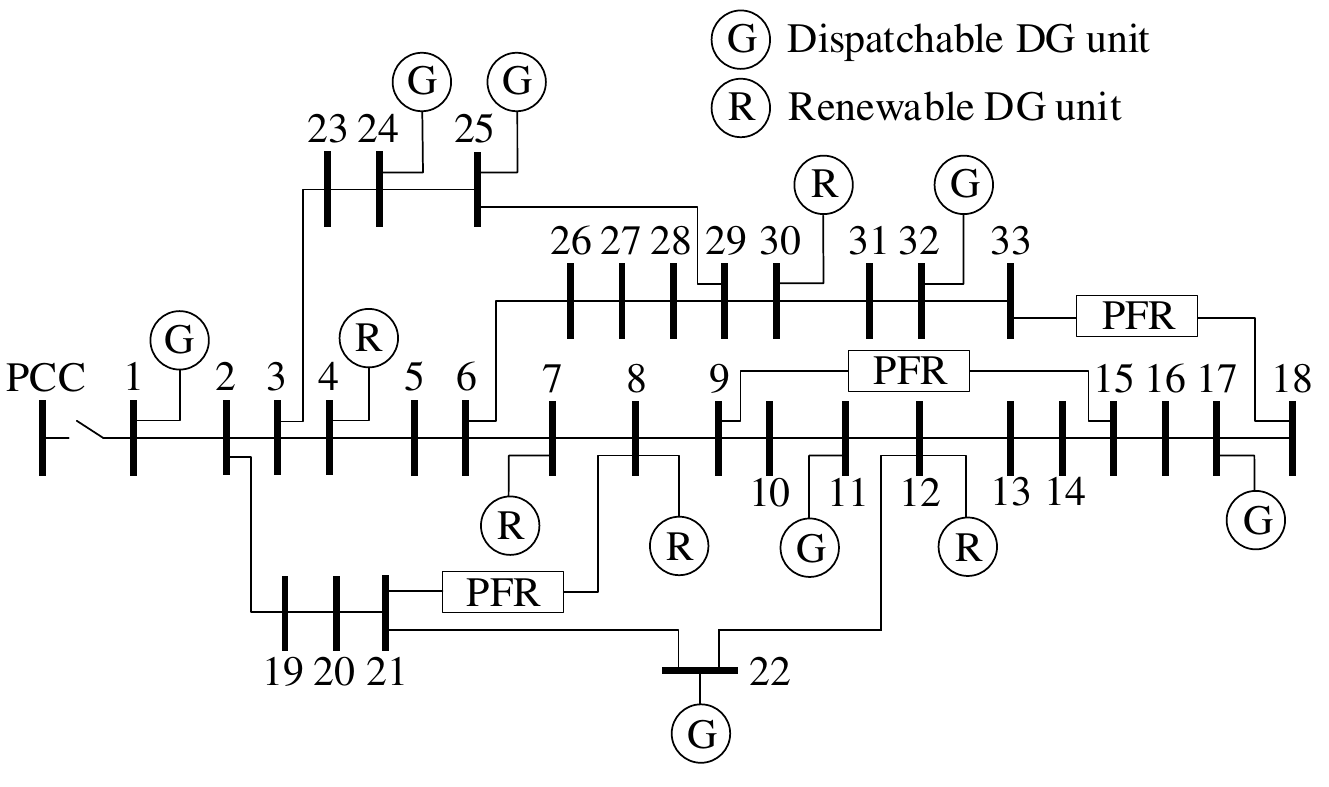}  \vspace{-0.3cm}
 \caption{The diagram of modified IEEE 33-bus microgrid.} \vspace{-0.3cm}
 \label{fig:test system}
\end{figure}

\begin{table}[!t] 
\renewcommand{\arraystretch}{1.3} 
\centering
\caption{Forecasted renewable-based DG outputs (MW)} 
\label{table:1}
 \begin{tabular}{ c| c| c| c| c| c }
\hline \hline
Bus  & 4 & 7 & 8 & 14 & 30    \\ 
  \hline
$P_W$  & 0.6& 0.2 & 0.5& 0.7& 0.4    \\   
\hline \hline
\end{tabular} \vspace{-0.3cm}
\end{table}  

\section{Case Study} \label{sec:simulation}
We use the modified IEEE 33-bus system shown in Fig. \ref{fig:test system} to test the performance of the proposed CC-OPF-PFR model and the solution method. The line parameters and the loads are the same as those in Matpower \cite{5491276}. PFRs are installed at lines (8,21), (9,15) and (18,33). Seven dispatchable DGs and five renewable DGs are installed in the system. Each renewable DG operates at the pre-defined power factor $\tan{\phi}_i = 0.95$. Their forecasted active power outputs are listed in Table \ref{table:1} and the covariance matrix is omitted here due to the space limit. Note that the total active power load of the system is 3.72 MW and the renewable penetration level is around 65$\%$.
Voltage limits for all the buses are set to ${V}_i^{\text{min}} = 0.95$ and ${V}_i^{\text{max}} = 1.05$, respectively. 
The PFR parameter specifications follow  \cite{chen2020robust}, particularly  ${\gamma}_{i_j}^{\text{min}} = 0.8, {\gamma}_{i_j}^{\text{max}} = 1.2, {\beta}_{i_j}^{\text{max}} =- {\beta}_{i_j}^{\text{min}} = 20^{\text{o}} $.  
The optimization computation is conducted on a 64-bit computer with 3.2 GHz CPU and 16 GB RAM. The optimization problem is solved by Mosek via CVX \cite{cvx} in Matlab.

For comparison, we obtain the optimal solutions from the following four versions of OPF: (a) Normal OPF without PFRs and renewable uncertainties; (b) OPF-PFR (i.e., OPF with PFRs and without renewable uncertainties); (c) CC-OPF without PFRs; (d) CC-OPF-PFR.

For the base case, we set the violation probabilities as $\epsilon = \epsilon_P=\epsilon_Q=\epsilon_V=\epsilon_{\omega} = 0.01$ and the tolerance value for convergence as $\delta = 10^{-5}$. Thus, the probability of satisfying the chance constraints is no less than $99\%$. The droop gains are set as $K_{Pi} = 3$ and $K_{Qi} =30$ for dispatchable DGs.

To verify the proposed method, we also calculate the empirical constraint violation probabilities and probability density functions (PDFs) of voltage magnitudes by Monte Carlo Simulation (MCS). In the MCS, we calculate the AC power flow under $10^4$ renewable power scenarios following the prescribed multivariate normal distribution.


\vspace{-0.3cm}
\subsection{Merits of CC-OPF-PFR}
We compare the generation costs, computational times, iteration numbers, and the maximum empirical violation probabilities Max.$\epsilon_{emp}$ of (a)--(d) and the results are listed in Table \ref{table:2}. The iterative algorithm normally converges in 3 iterations and the computational time for two CC-OPF (c)--(d) problems are both within 10 seconds. This highlights the efficiency of the proposed iterative algorithm.
For the optimal solutions given by normal OPF and OPF-PFR, the maximum empirical violation probabilities are over 50$\%$, which indicates the deterministic OPF is not sufficient to maintain an acceptable voltage profile under renewable uncertainties. By comparison, CC-OPF and CC-OPF-PFR control the maximum empirical violation probabilities below the pre-specified level (1$\%$). This result verifies the effectiveness of chance constraints in securing the droop-controlled MGs against renewable uncertainties. Moreover, the cost of CC-OPF-PFR is $2.0\%$ lower than CC-OPF and even sightly lower than normal OPF. It shows that PFRs help the system in achieving better security without deteriorating the operational economy.

\begin{table}[t] 
\renewcommand{\arraystretch}{1.3} 
\centering
\begin{threeparttable}
\caption{Results of the Four OPF Approaches}
\label{table:2}
 \begin{tabular}{  c c c c c  }
 \hline \hline
Methods & (a)  & (b)  & (c) & (d) \\ 
  \hline
Cost (\$/hr) &2898.31  & 2836.89 & 2920.62  & 2841.63  \\ 
  \hline
CPU time (s) 		& 1.04 &  3.33  & 4.04 & 9.94   \\  \hline
Iterations 		&/ &  /  & 3 & 3  \\  \hline
Max.$\epsilon_{emp}^*$  	& 52.46\% & 58.09\%      &0.83\%	    &0.14\%    \\
\hline \hline
\end{tabular} 
\begin{tablenotes}
      \footnotesize
      \item *Max.$\epsilon_{emp}$ refers to the maximum empirical violation probability for all the constraints.
    \end{tablenotes}
\end{threeparttable} \vspace{-0.5cm}
\end{table}

\begin{figure}[t]
\includegraphics[width=\linewidth]{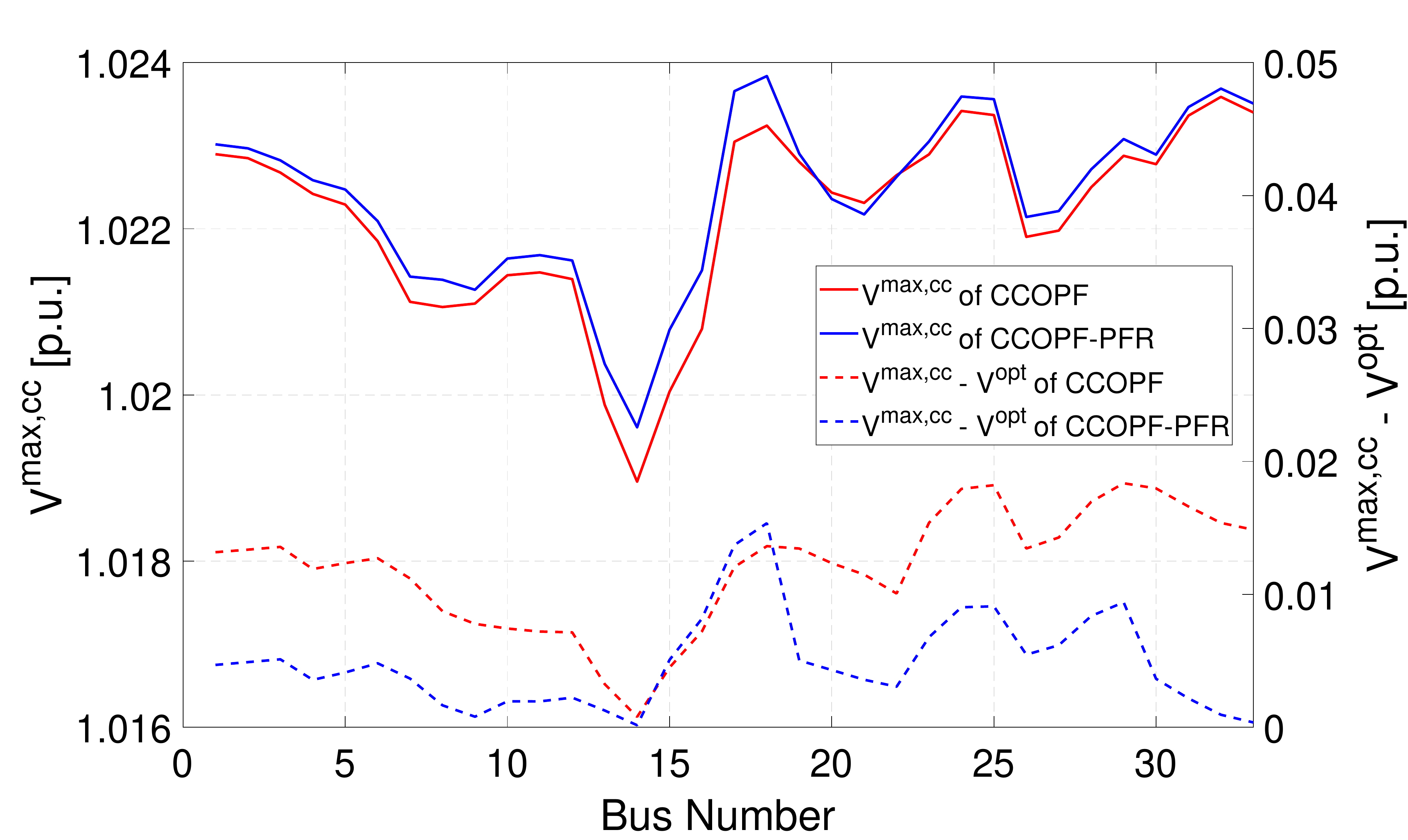} \vspace{-0.7cm}
\caption{$\bm{V}^{\text{max,cc}} - \bm{V}^{\text{opt}}$ and equivalent voltage limits $\bm{V}^{\text{max,cc}}$ in $ \eqref{CCOPF-r}$ } \vspace{-0.4cm}
\label{fig: Voltage margin}
\end{figure}

\begin{figure}[t]
\includegraphics[width=\linewidth]{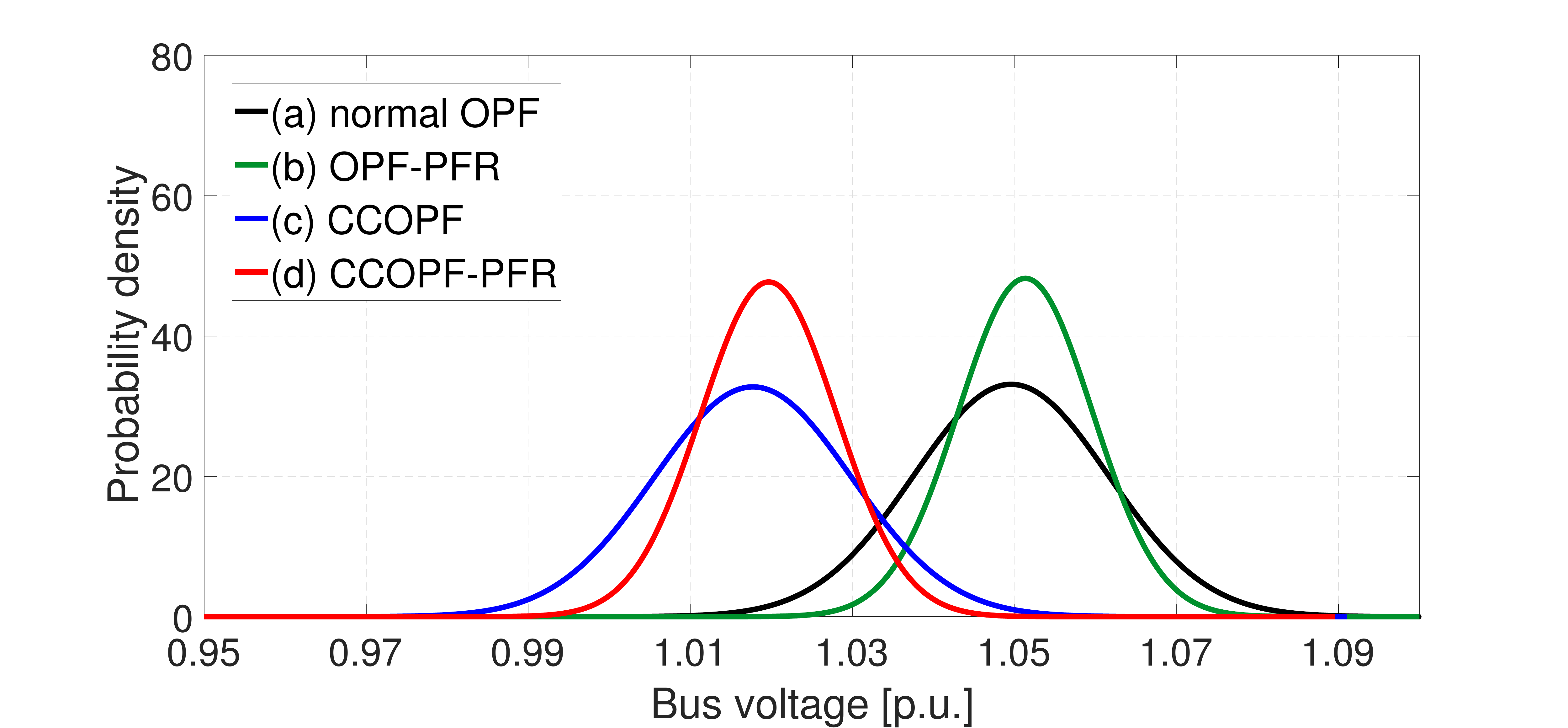} 
\vspace{-0.7cm}
\caption{Empirical voltage PDFs at bus 14 under (a)-(d)($K_{Pi} =3, K_{Qi} = 30$)} \vspace{-0.5cm}
\label{fig:Voltage 1}
\end{figure}

To detail the contribution of PFRs, we define $V_i^{\text{max,cc}} = V_i^{\text{max}} - \Omega_{V_i}$ as the equivalent voltage upper limit in voltage constraint \eqref{LCC-5} (i.e., $V_i(\bm{0})  \leq V_i^{\text{max,cc}}, \forall i\in \mathcal{N}$) when the iterative algorithm converges. Also, we denote $\bm{V}^{\text{opt}}$ as the voltage magnitudes at the optimal solution. Fig. \ref{fig: Voltage margin} shows the values of $\bm{V}^{\text{max,cc}}$ and $\bm{V}^{\text{max,cc}} - \bm{V}^{\text{opt}}$ obtained by CC-OPF and CC-OPF-PFR, respectively. 
At most buses, the values of $\bm{V}^{\text{max,cc}}$ under CC-OPF-PFR are higher than that under CC-OPF. It means that the voltage standard deviations of most buses are reduced by tuning $\bm{T}^*$ and $\bm{\beta}^*$ of the PFRs. As a result, the feasible region of the optimization problem is also enlarged. In particular, we refer to bus 14 as the critical bus because the voltage magnitude at bus 14 hits the equivalent voltage limit (i.e., $V_{14}^{\text{max,cc}} - V_{14}^{\text{opt}} = 0$) under both CC-OPF and CC-OPF-PFR. The hitting under CC-OPF (see the red dashed line) prevents the problem from seeking a further better solution along this direction; the hitting under CC-OPF-PFR (see the blue dashed line) means CC-OPF-PFR allows for pursuing a better solution due to the enlarged feasible region and this solution can be further improved by incorporating more network flexibility, i.e., more PFRs.  


 

To further illustrate the effects of PFRs, the empirical voltage PDFs at bus 14 under (a)--(d) are shown in Fig. \ref{fig:Voltage 1}.
By comparing (a) with (b) and (c) with (d), we observe that the empirical voltage PDFs have significantly narrower shapes if PFRs are included in the system. This is consistent with our previous analysis in Section \ref{sec:method} that tuning  the PFR parameters $\bm{T}^*$ and $\bm{\beta}^*$ is effective in reducing the voltage standard deviations. Thus, the voltage volatility levels under OPF-PFR and CC-OPF-PFR are both lower than that without PFRs. Moreover, by comparing (a) with (c) and (b) with (d), we observe that the inclusion of chance constraints lead to left shifts of the mean values which are mainly achieved by power injection changes. On the other hand, the power injections do not make significant contribution to voltage volatility reduction as the PDF curves in (a) and (c) have almost the same shape. 
From the above discussion, we reveal that power injections and PFRs have different mechanisms in voltage regulation under uncertainties. Power injections mainly contribute to the change of voltage mean values, while PFRs mainly contribute to the reduction of voltage variance. Therefore, the proposed CC-OPF-PFR outperforms the traditional CC-OPF by introducing a new dimension of control mechanism.

\vspace{-0.3cm}
\subsection{Performance under Different Droop Gains}

%
%

To further illustrate the merits of the CC-OPF-PFR model, we compare its performance with CC-OPF under another two types of droop gains in terms of generation costs, and voltage volatility. We set $K_{Pi} = 1, K_{Qi} =10$ referring to low droop gains and $K_{Pi} = 5, K_{Qi} =50$ referring to high droop gains.

1) Generation cost. 
For CC-OPF, the generation costs are 2898.89 $\$$/hr under low droop gains and 3007.42 $\$$/hr under high droop gains. For CC-OPF-PFR, the generation costs are 2838.94 $\$$/hr under low droop gains and 2849.33 $\$$/hr under high droop gains. 
We observe that higher droop gains always lead to higher generation cost and this kind of cost increase is more notable for CC-OPF because it relies on tuning the power injections only. In addition, CC-OPF-PFR introduces 2.07$\%$ and 5.25$\%$ cost reduction compared to CC-OPF under low and high droop gains, respectively. This again highlights the economic merits brought by PFRs and this value is more remarkable under higher droop gains. 

\begin{figure}[t]
 \includegraphics[width=\linewidth]{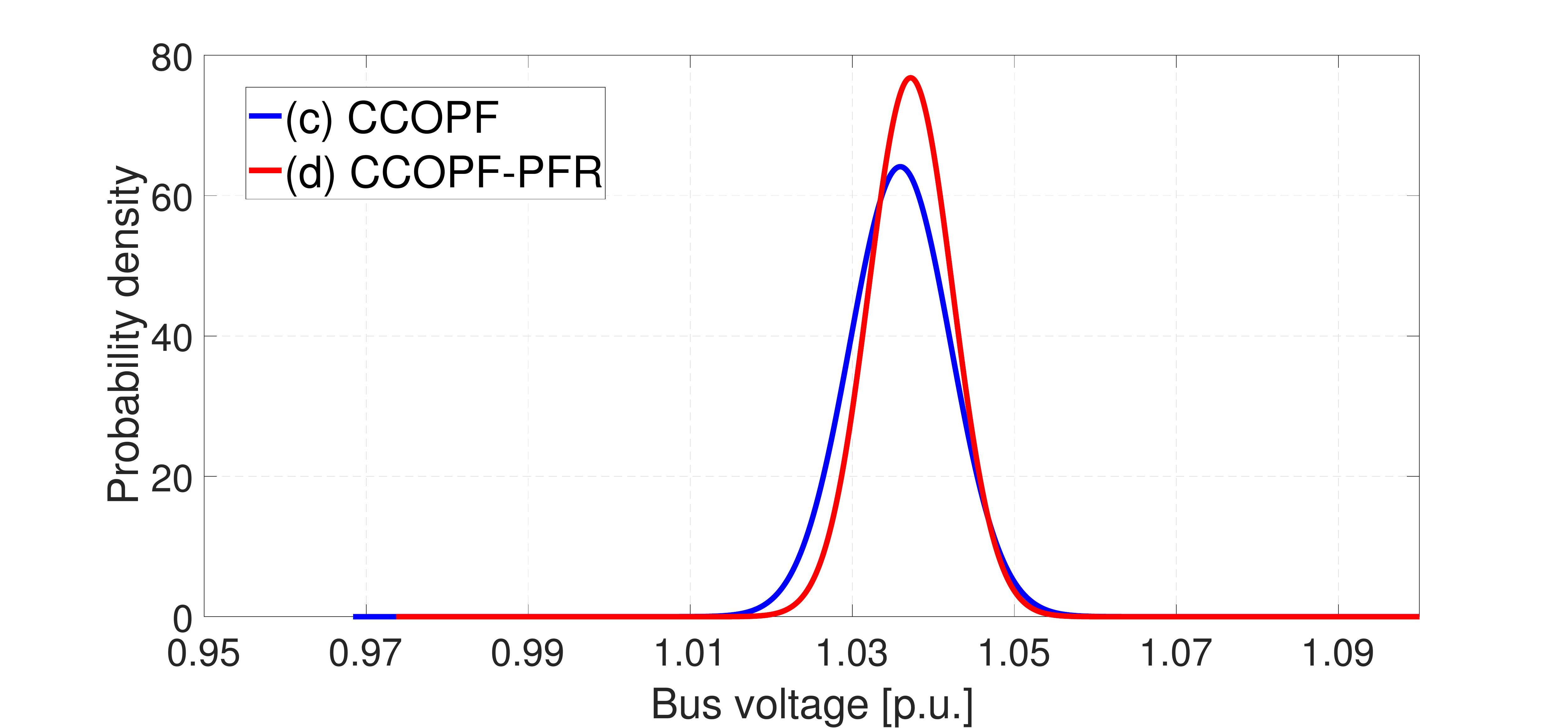}
 \vspace{-0.5cm} 
 \caption{Empirical voltage PDFs at bus 14 under (c)(d) ($K_{Pi} =1, K_{Qi} = 10$)} \vspace{-0.5cm}
 \label{fig:DP10}
\end{figure}

\begin{figure}[t]
 \includegraphics[width=\linewidth]{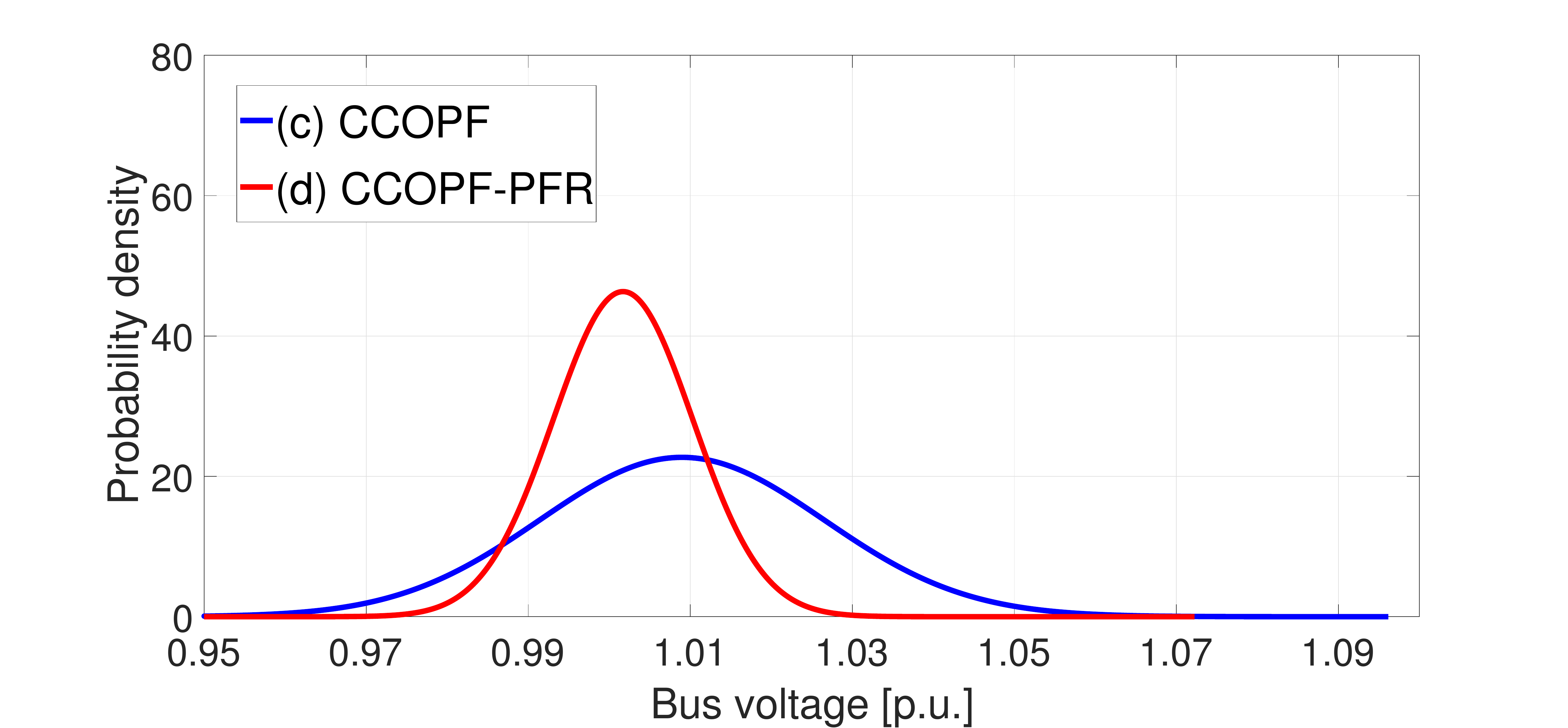}
 \vspace{-0.5cm} 
 \caption{Empirical voltage PDFs at bus 14 under (c)(d) ($K_{Pi} =5, K_{Qi} = 50$)} \vspace{-0.5cm}
 \label{fig:DP50}
\end{figure}

2) Voltage volatility. Similar to the previous subsection, Fig. \ref{fig:DP10} and Fig. \ref{fig:DP50} show the empirical voltage PDFs of bus 14 under low and high droop gains.  We observe that the voltage volatilities are low for both CC-OPF and CC-OPF-PFR under low droop gains. However, under high droop gains, CC-OPF introduces a rather high voltage volatility level while CC-OPF-PFR still keeps a relatively small voltage volatility level.

We further discuss the results as follows.
\begin{enumerate}[(a)]
\item Under low droop gains, the voltage standard deviations $\text{Dev}\{\bm{V}(\bm{\xi})\}$ are naturally small. This property leads to small voltage margins $\bm{\Omega}_V$ and the chance constraints can be easily satisfied without PFRs. Even with PFRs, their capabilities on $\text{Dev}\{\bm{V}(\bm{\xi})\}$ reduction by tuning $\bm{T}^*$ and $\bm{\beta}^*$ is limited to a small range. Therefore, the voltage volatility levels under CC-OPF and CC-OPF-PFR are similarly low as shown in Fig. \ref{fig:DP10}.
\item Under high droop gains, the voltage standard deviations $\text{Dev}\{\bm{V}(\bm{\xi})\}$ are relatively large and thus lead to more stringent voltage constraints \eqref{LCC-5}--\eqref{LCC-6}. In this case, the value of PFRs is also amplified. Different from the CC-OPF which can only adjust the power injections to satisfy the constraints, PFRs can meet the voltage constraints by tuning $\bm{T}^*$ and $\bm{\beta}^*$ for smaller voltage standard deviations. 
Therefore, the voltage volatility under CC-OPF-PFR is much lower than that under CC-OPF as shown in Fig. \ref{fig:DP50}. Furthermore, with the help of PFRs, the voltage volatility levels under high droop gains can even be close to the case under low droop gains.
\end{enumerate}

From the above discussion, we conclude that if high droop gains are adopted for islanded MGs, it could lead to serious volatile voltages under traditional CC-OPF. But the high droop gains also amplify the capability of PFRs in reducing  cost and voltage standard deviations. Therefore, the proposed CC-OPF-PFR has more significant merits under higher droop gains, which well matches the high droop gain nature of MGs.

\vspace{-0.2cm}
\section{Conclusion} \label{sec:conclusion}
In this paper, we propose a new CC-OPF-PFR problem in droop-controlled MGs under renewable uncertainties. In this formulation, the droop characteristics and PFRs are for the first time both considered in the CC-OPF model, and the PFRs provide a new network-oriented mechanism different form the traditional node-side control. 
Due to the complexity of AC power flow with PFRs, we establish an iterative solution method for the CC-OPF-PFR. Using sensitivity analysis, the subproblem in each iteration is transformed into a deterministic optimization problem that can be efficiently solved by existing solvers such as SDP relaxation, which provides tractability and efficiency. Numerical results show that the proposed CC-OPF-PFR model significantly reduce the voltage volatility under high droop gains and achieves a highly secure solution with lower cost than the case without PFRs. 

Many directions can be considered for future research, e.g., data-driven methods can be included for constructing the ambiguity set and distributionally robust optimization can be developed. Security and stability constraints need to be considered for a more robust and secure operating point.

\bibliographystyle{IEEEtran}
\vspace{-0.3cm}
\bibliography{IEEEabrv,Reference}

\end{document}